\documentclass[apj]{emulateapj}
\shorttitle{Constraints on UHECR sources}
\shortauthors{Waxman \& Loeb}

\begin{document}

\title{Constraints on the Local Sources of Ultra High-Energy Cosmic Rays}

\author{Eli Waxman\altaffilmark{1} \& Abraham
Loeb\altaffilmark{2}$^{,}$\altaffilmark{3}}
\altaffiltext{1}{Physics Faculty,
Weizmann Institute of Science, Rehovot 76100, Israel}
\altaffiltext{2}{Astronomy Department, Harvard University, 60 Garden
Street, Cambridge, MA 02138, USA}
\altaffiltext{3}{Einstein Minerva center,
Weizmann Institute of Science, Rehovot 76100, Israel}

\begin{abstract}

Ultra high-energy cosmic rays (UHECRs) are believed to be protons
accelerated in magnetized plasma outflows of extra-Galactic sources.
The acceleration of protons to $\sim10^{20}$~eV requires a source
power $L>10^{47}~{\rm erg~s^{-1}}$.  The absence of steady sources of
sufficient power within the GZK horizon of 100~Mpc, implies that UHECR
sources are transient.  We show that UHECR "flares" should be
accompanied by strong X-ray and $\gamma$-ray emission, and that X-ray
and $\gamma$-ray surveys constrain flares which last less than a
decade to satisfy at least one of the following conditions: {\it (i)}
$L>10^{50}{\rm erg~s^{-1}}$; {\it (ii)} the power carried by
accelerated electrons is lower by a factor $>10^2$ than the power
carried by magnetic fields or by $>10^3$ than the power in accelerated
protons; or {\it (iii)} the sources exist only at low redshifts,
$z\ll1$. The implausibility of requirements {\it (ii)} and {\it (iii)}
argue in favor of transient sources with $L>10^{50}{\rm erg~s^{-1}}$.

\end{abstract}

\keywords{cosmic-rays --- X-rays: general --- gamma rays: observations
--- galaxies: nuclei}

\section{Introduction}
\label{sec:intro}

The origin of ultra high-energy ($>10^{19}$~eV) cosmic-rays (UHECRs)
remains a mystery \citep{Bhattacharjee:1998qc,Nagano:2000ve}.  The
sources have not been robustly identified, and the models of particle
acceleration are challenged by the fact that the energy spectrum
extends to $>10^{20}$~eV. Several observational clues suggest that the
UHECR flux is dominated by extra-Galactic light nuclei: the spectrum
flattens at $\sim10^{19}$~eV \citep{Nagano:2000ve}, there is evidence
for a composition change from heavy to light nuclei at
$\sim10^{19}$~eV \citep{Bird93,Abbasi05}, and the UHECR arrival
direction distribution is nearly isotropic \citep{Finley04,HiResIso}.
The recent detection of a weak anisotropy in the arrival distribution
of $>6\times10^{19}$~eV cosmic-rays \citep{auger}, is consistent with
that predicted by assuming that the spatial distribution of UHECR
sources correlates with the large-scale distribution of galaxies
\citep{Fisher97,Kashti08}.

Although the identity of the UHECR particles is uncertain, we will
assume here that they are protons. This assumption is motivated by two
arguments. First, the observed spectrum of cosmic-rays with energies
$>10^{19}$~eV is consistent with a cosmological distribution of proton
accelerators producing (intrinsically) a power-law spectrum of high
energy protons with $d\log N/d\log E\approx -2$, for the number $N$ as
a function of energy $E$ \citep{Waxman95,Bahcall03,Kashti08}.  This
intrinsic power-law spectrum is consistent with that expected in
models of particle acceleration in collisionless shocks, for both
non-relativistic~\citep{Blandford87} and relativistic shocks (Waxman
2006; see however Keshet 2006).  Second, the leading candidates for
extra-Galactic sources, namely gamma-ray bursts and active galactic
nuclei (see below), are expected to accelerate primarily protons.

Robust model-independent considerations imply that UHECR protons can
only be produced by sources with an exceedingly high power output
\citep{Waxman_CR_rev}, $L>\Gamma^2\beta^{-1}10^{46}~{\rm erg~s^{-1}}$,
where $\Gamma$ and $\beta c$ are the Lorentz factor and characteristic
velocity associated with plasma motions within the
source\footnote{Somewhat more stringent limits may be obtained by
specifying the acceleration process; see \citet{Norman95},
\citet{Waxman95prl}, and \S~\ref{sec:rates} below.}. Since no steady
source above this power threshold is known to exist within the 100~Mpc
{\it GZK horizon}, the distance to which the propagation of
$\sim10^{20}$~eV protons is limited by their interaction with the
cosmic microwave background \citep{G66,ZK}, the UHECR sources are most
likely transient. A possible alternative is, of course, an unknown
class of "dark sources", which produce little radiation and therefore
remain undetectable by telescopes.

Only two types of sources are known to satisfy the above minimum power
requirement: active galactic nuclei (AGN) -- the brightest known
steady sources, and gamma-ray bursts (GRBs) -- the brightest known
transient sources\footnote{It was recognized early on (\cite{Hillas}
and references therein) that while highly magnetized neutron stars may
also satisfy the minimum power requirement, it is difficult to utilize
the potential drop in their electro-magnetic winds for proton
acceleration to ultra-high energy (see, however, \cite{Arons}).}. The
absence of AGN with $L>10^{46}~{\rm erg~s^{-1}}$ within the GZK
horizon had motivated \citet{Gruzinov} to suggest that UHECRs may be
produced by a new, yet undetected, class of short duration AGN flares
resulting from the tidal disruption of stars or accretion disk
instabilities.

We show in \S~\ref{sec:flares} that if electrons are accelerated
together with the protons in UHECR-producing flares, then their
radiative losses will produce a bright flare of X-ray and $\gamma$-ray
photons. We then show in \S~\ref{sec:LFs} that existing X-ray
and $\gamma$-ray surveys already put stringent constraints on the
properties of UHECR flares. In \S~\ref{sec:hidden} we discuss the
possibility of "hiding" the X-ray emission. Our conclusions are
summarized in \S~\ref{sec:conclusions}.

Throughout our discussion, we consider a scenario in which the flare
is associated with ejection of magnetized plasma from the source, and
where the charged particles are accelerated within the magnetized
outflow. The non thermal emission from a wide range of sources is
described within the framework of such a scenario. This includes AGN
jets and GRBs, as well as the transient AGN flares proposed by
\citet{Gruzinov}. We parametrize the UHECR flares by their power, $L$,
duration, $\Delta t$, characteristic ejection speed $\beta c$, and
rate per unit volume in the local Universe, $\dot{n}$. The fractions
of the total energy output carried by protons, electrons and magnetic
fields are denoted by $\epsilon_p$, $\epsilon_e$ and $\epsilon_B$,
respectively, and we assume that the energy spectrum of accelerated
electrons is similar to that of accelerated protons with a power-law
index, $d\log N/d\log E\approx -2$.  This index is expected for
astrophysical sources which accelerate particles in strong
collisionless shocks \citep{Blandford87,Waxman_CLS_rev}, and is
inferred from the radiation observed from a variety of high energy
sources, such as supernova remnants \citep{SN} and GRBs
\citep{WAG97a}.

It should be pointed out that the composition of the jets of high
energy sources is unknown. Two classes of models are generally
discussed: jets where the energy flux is dominated by the plasma
kinetic energy, and jets where most of the energy is carried by
electromagnetic flux. For the "kinetic" jets, a plausible mechanism
exists for energy dissipation, particle acceleration and radiation
emission, namely internal collisionless shocks within the
outflow. Within this mechanism, a particle distribution following
$d\log N/d\log E\approx -2$ is naturally expected. For the
"electromagnetic" jets, the mechanism of energy dissipation and
particle acceleration is not well understood. We will assume that a
$d\log N/d\log E\approx -2$ particle distribution is generated in this
model too, as suggested by observations.

As explained at the end of \S~\ref{sec:rates}, our conclusions are
valid for both spherical and jetted flows.  Throughout the paper,
$L$ stands for the "isotropic-equivalent power" (i.e. for a flow which
is conical rather than spherical, $L$ stands for the power that would
have been carried by the flow had it been spherically symmetric), and
$\dot{n}$ stands for the "isotropic-equivalent rate density" (i.e. the
rate inferred under the assumption of spherically symmetric emission).

\section{Flare properties}
\label{sec:flares}

\subsection{Rates and luminosities}
\label{sec:rates}

We define a transient source to have an active phase of duration
$\Delta t$ shorter than the time delay $\Delta t_{CR}$ between the
photon and the UHECR arrival times. With this definition, a ``steady''
source is one which is still active when the UHECRs from it are
being detected.  The arrival time delay originates from deflections of the
charged UHECRs by intergalactic magnetic fields, and can be expressed
in terms of the deflection angle, $\theta$, and propagation distance,
$d$, as $\Delta t_{CR}\approx \theta^2 d/4c$. The deflection angle is
limited to $\lesssim 1^\circ (d/100~{\rm Mpc})^{1/2}(E/10^{20}~{\rm
eV})^{-1}$ \citep[see detailed discussion in \S 2.2 of][]{Kashti08},
and therefore
\begin{equation}\label{eq:dtCR}
    \Delta t_{CR}\lesssim 10^{4.5}(d/100~{\rm Mpc})^2(E/10^{20}~{\rm
eV})^{-2}\,\rm yr.
\end{equation}
For transient sources, the apparent number density of UHECR sources is
energy dependent and given by $\dot{n}\Delta t_{CR}$.

The required number density of active flares is obtained from the observed
energy production rate of UHECR protons per comoving volume,
$\dot{\varepsilon}\equiv E^2 d\dot{n}_p/dE=0.7\pm0.3\times10^{44}{\rm
erg~Mpc^{-3}~yr^{-1}}$ \citep{Waxman95,Bahcall03}, giving
\begin{equation}\label{eq:n_flares}
   \dot{n}\Delta t=\frac{\dot{\varepsilon}}{\epsilon_p L/\Lambda}
                   =3.2\times10^{-10}\frac{\dot{\varepsilon}_{44}\Lambda_1}{\epsilon_p
                   L_{47}} {\rm Mpc}^{-3},
\end{equation}
where, $\dot{\varepsilon}_{44}\equiv \dot{\varepsilon}/10^{44}{\rm
erg~Mpc^{-3}~yr^{-1}}$, $L_{47}\equiv (L/10^{47}{\rm erg~s^{-1}})$,
$\epsilon_pL$ is the total energy output in protons, and
$\epsilon_pL/\Lambda$ is the energy production per logarithmic proton
energy ($E$) interval in the observed energy range. For the acceleration
spectrum of strong collisionless shocks, $E^2 d\dot{n}_p/dE \sim const$, we
get $\Lambda=\ln(E_{\rm max}/E_{\rm min})$ and so
$\Lambda_1\equiv(\Lambda/10)\sim 1$.

The flare duration is limited by the absence of UHECR sources with
multiple events, the so-called `repeaters'. The absence of repeaters
sets a lower limit on the number density of sources, $n$
\citep{Fisher97,Kashti08}.  This can be derived by noting that the
nearly isotropic distribution of the $\sim30$ {\it Auger} events of
energy $>6\times10^{19}$~eV requires tens of sources to be active
within $\approx200$~Mpc (the propagation distance of protons with
energies $>6\times10^{19}$~eV) within {\it Auger}'s field-of-view,
implying $n\gtrsim10^{-5.5}{\rm Mpc}^{-3}$.  For transient sources
this requirement implies $\dot{n}\Delta t\gtrsim10^{-5.5} (\Delta
t/\Delta t_{CR}){\rm Mpc}^{-3}$. For a GZK horizon distance of
$200~{\rm Mpc}$ corresponding to UHECR energy of $E=6\times10^{19}{\rm
eV}$, we get $\Delta t_{CR}\lesssim 10^{5.5}$~yr, and therefore
\begin{equation}\label{eq:n_lim}
    \dot{n}\Delta t\gtrsim 3\times10^{-10}n_{-5}
\left(\frac{\Delta t}{10\,\rm yr}\right)\rm Mpc^{-3},
\end{equation}
where $n_{-5}\equiv (n/10^{-5}{\rm Mpc}^{-3})$. Using
Eq.~(\ref{eq:n_flares}) we find
\begin{equation}\label{eq:dt_lim}
    \Delta t\lesssim10 n_{-5}^{-1}\frac{
\dot{\varepsilon}_{44}\Lambda_1}{\epsilon_p L_{47}} \,{\rm yr}.
\label{dt}
\end{equation}

The number density of associated photon flares may be obtained by
assuming that: {\it (i)} the accelerated electrons have the same
initial power-law index for their energy spectrum as the protons, and
{\it (ii)} the electrons lose all their energy to radiation (see \S
\ref{sec:e_loss} below for the justification of the latter
assumption). The photon luminosity per logarithmic frequency ($\nu$)
interval is then $\nu L_\nu=\epsilon_e
L/2\Lambda=(\epsilon_e/2\epsilon_p)(\epsilon_p L/\Lambda)$, where the
factor of 2 is introduced since typically $\nu\propto E_e^2$. This
implies that the number density of active photon flares with a
luminosity $\gtrsim \nu L_\nu$ is
\begin{equation}\label{eq:n_phot_flares}
    \dot{n}\Delta t=\frac{\epsilon_e}{2\epsilon_p}\frac{\dot{\varepsilon}}
    {\nu L_\nu} =1.6\times10^{-10}\frac{\epsilon_e}{\epsilon_p}
    \frac{\dot{\varepsilon}_{44}}{(\nu L_\nu)_{46}} {\rm Mpc}^{-3},
\end{equation}
where $(\nu L_\nu)_{46}\equiv (\nu L_\nu/10^{46}{\rm erg~s^{-1}})$.

Requiring that the acceleration time $t_{\rm acc}$ be smaller than the
plasma expansion time $t_{\rm dyn}$ and the proton energy loss time
$t_{\rm loss}$, sets lower limits on $L$ and the outflow Lorentz
factor, $\Gamma$ \citep{Waxman95prl}. In the following, we briefly
describe these limits and derive the implied constraints on the photon
luminosity. Assuming that acceleration results from electromagnetic
processes within the outflowing plasma, the acceleration time must
exceed the Larmor gyration time of the accelerated
particle\footnote{For acceleration in collisionless shocks, $t_{\rm
acc}$ is larger than the Larmor time by a factor $\sim (c/v)^2$ where
$v$ is the shock velocity in the plasma rest frame.}, $t_{\rm
acc}\gtrsim 2\pi f R_L/c=2\pi f E'/eBc$, where $E'=E/\Gamma$ and $f$
is a dimensionless factor of order a few which depends on the details
of the acceleration mechanism, and where the various times are defined
in the plasma rest frame.  Requiring $t_{\rm acc}<t_{\rm dyn}=
r/\Gamma\beta c$, where $r$ is the radial distance from the source at
which particle acceleration takes place, this implies $B>f E/\beta er$
and equivalently, \citep{Waxman95prl}
\begin{equation}\label{eq:L_Bmin}
    \epsilon_B L>2\left(\frac{\pi f\Gamma E}{e}\right)^2c
    =6.6\times10^{46}f^2\frac{\Gamma^2}{\beta}E_{20}^2\,{\rm erg~s^{-1}},
\end{equation}
where $\epsilon_B L=4\pi r^2 c \Gamma^2 B^2/8\pi$,
and $E_{20}=(E/10^{20}$~eV). The minimum photon luminosity is therefore
\begin{eqnarray}\label{eq:Lmin}
    \nu L_\nu&>&3.3\times10^{45}\frac{f^2\epsilon_e}{\Lambda_1\epsilon_B}
    \frac{\Gamma^2}{\beta}E_{20}^2\,{\rm erg~s^{-1}}\cr
    &>&8.6\times10^{45}\frac{f^2\epsilon_e}{\Lambda_1\epsilon_B}
    E_{20}^2\,{\rm erg~s^{-1}}.
\end{eqnarray}
One of our primary objectives is to demonstrate that
exceptionally powerful flares with $L>10^{50}{\rm
erg~s^{-1}}$ are required. In what follows we limit the
discussion to flares with $\Gamma<10^{1.5}$, since a higher $\Gamma$ implies
$L>10^{50}\epsilon_B^{-1}{\rm erg~s^{-1}}$.

A lower limit on the bulk Lorentz factor $\Gamma$ is set by requiring
that the synchrotron loss time would exceed the acceleration time,
$t_{\rm acc}< t_{\rm loss}=6\pi\Gamma(m_pc^2)^2
/cB^2\sigma_T(m_e/m_p)^2E$, where $m_p$ and $m_e$ are the proton and
electron masses.  Using $r<2\Gamma^2 c \Delta t$, this condition
implies \citep{Waxman95prl}
\begin{equation}
\beta^{1/2}\Gamma>
\left({f\sigma_T\over 6\pi e}\right)^{1/5} \left({m_e\over m_p}{E\over
m_pc^2}\right)^{2/5} \left({\epsilon_B
L\over 2c^3}\right)^{1/10}\Delta t^{-1/5},
\end{equation}
or numerically,
\begin{equation}\label{eq:Gamma}
    \Gamma>1.1 f^{1/5}E_{20}^{2/5}(\epsilon_B L_{47})^{1/10}\Delta t_{\rm
    yr}^{-1/5}.
\end{equation}
Hereafter, we drop the dependence on $\beta$ since the flow is required to
be at least mildly relativistic with $\beta\approx1$. Combining this result
with Eq.~(\ref{eq:L_Bmin}), we get
\begin{equation}\label{eq:L_Bmin1}
       \epsilon_B L>7.9\times10^{46}f^3E_{20}^{7/2}\Delta t_{\rm
       yr}^{-1/2}\,{\rm erg~s^{-1}}.
\end{equation}

The constraints on $L$ and $\Gamma$ are the same for a spherical and
conical (jet-like) outflow as long as the opening angle of the jet
$\theta_j$ is larger than $1/\Gamma$ \citep{Waxman95prl}. Thus, the
constraints in equations~(\ref{eq:L_Bmin})--(\ref{eq:L_Bmin1}) apply
in both cases, provided that $L$ is interpreted as the
isotropic-equivalent power.  In the case of jets, there could be a
discrepancy between the apparent number of UHECR sources and photon
sources, in case the deflection angle of CRs by the intergalactic
magnetic field is larger than $\max[\theta_j,1/\Gamma]$. For the
Lorentz factors considered here, $\Gamma<10^{1.5}$, the magnetic
deflections are smaller than $1/\Gamma>1/30=2^\circ$ (see the opening
paragraph of this section), implying that we should see the same
sources in both photons and UHECRs. Under these circumstances, the
results in equations ~(\ref{eq:n_lim})--(\ref{eq:n_phot_flares}) hold,
provided that $n$ and $\dot{n}$ refer to the isotropic equivalents
quantities. This also implies that we can use isotropic equivalent
luminosities in the luminosity functions.

\subsection{High energy photon spectrum}
\label{sec:e_loss}

We next show that our estimate of the photon luminosity, $\nu
L_\nu=\epsilon_e L/\Lambda$, is valid for photon energies above $\sim
1$~keV.  We start our discussion by justifying the fast cooling
assumption for the electrons.

If electrons are accelerated through a process similar to that of the
protons, their acceleration time to a given energy would be similar to
that of the protons.  The maximum electron energy would be lower than
that of the protons, owing to their higher cooling rate. Equating the
electron acceleration time to the radiative cooling time, $2\pi f
\gamma_e m_e c^2/eBc \sim m_e c^2/\sigma_T \gamma_e c u$, yields
$\gamma_e^2\sim eB/2\pi f \sigma_T u$. Here $\gamma_e$ is the electron
Lorentz factor in the plasma rest frame and $u$ is the electromagnetic
energy density of the radiation together with the magnetic fields. The
synchrotron photons emitted by the highest energy electrons carry an
energy, $\Gamma \hbar (0.3\gamma_e^2 eB/m_ec)=0.3\Gamma(\hbar e^2/2\pi
f \sigma_T m_e c ) B^2/ u$. Assuming that the electrons lose most of
their energy to radiation,
\begin{equation}\label{eq:nu_max}
    h\nu_{\rm syn,max}\approx8\Gamma\frac{\epsilon_B}{f(\epsilon_e+\epsilon_B)}\,{\rm MeV}.
\end{equation}
In the plasma rest frame, the electron Lorentz factor, $\gamma_c$, at which
the dynamical time is comparable to the synchrotron cooling time, is given
by $6\pi m_ec^2/\sigma_T c \gamma_c B^2=r/\Gamma c$. Approximating
$r=2\Gamma^2c\delta t$,
where $\delta t<\Delta t$ is the characteristic
variability time within the flare, we get
$\gamma_c=6\pi\Gamma^5\delta t m_ec^4/\sigma_T \epsilon_B L$. The
observed energy of the synchrotron photons emitted by electrons with a
Lorentz factor $\gamma_c$ is
$h\nu_c=0.3\hbar\Gamma\gamma_c^2eB/m_ec=5.4\sqrt{2}\pi^2e\hbar
c^{11/2}m_e\delta t\Gamma^8/\sigma_T^2(\epsilon_B L)^{3/2}$, namely
\begin{equation}\label{eq:nuc}
    \nu_{\rm syn,c}=0.1\Gamma^8(\epsilon_B L_{47})^{-3/2}\delta t_{\rm
    yr}\,{\rm GHz}.
\end{equation}
Using Eq.~(\ref{eq:L_Bmin}) we obtain
\begin{equation}\label{eq:nuc_max}
    \nu_{\rm syn,c}<0.1f^{-3}\Gamma^5E_{20}^{-3}\delta t_{\rm yr}\,{\rm
    GHz}.
\end{equation}

Equations~(\ref{eq:nuc_max}) and~(\ref{eq:nu_max}) imply that the
electrons lose most of their energy to radiation and that synchrotron
emission would lead to a flat spectrum, $\nu L_\nu=\epsilon_e
L/\Lambda$, from $\nu_{\rm syn,max}$ down to the larger frequency
among $\nu_{\rm syn,c}$ and $\nu_{\rm syn, min}$, the characteristic
synchrotron frequency of photons emitted by the lowest energy to which
electrons are accelerated.  The value of $h\nu_{\rm syn,min}$ depends
on the details of the acceleration mechanism.  As we show below,
$h\nu_{\rm syn,min}\ll1$~eV for acceleration in internal shocks within
an expanding wind, and a flat spectrum, $\nu L_\nu=\epsilon_e
L/\Lambda$, is expected down to optical frequencies. However, the
X-ray luminosity is expected to be comparable to $\epsilon_e
L/\Lambda$ also for $h\nu_{\rm syn,min}$ well above the X-ray band,
owing to electron cooling. Consider the extreme case where all electrons
are accelerated to the maximum energy. In this case the luminosity is
$\nu L_\nu=\epsilon_e L$ at $h\nu_{\rm syn,max}$, and the cooling
electrons would produce a luminosity $\nu L_\nu/\epsilon_e L\approx
(\nu/\nu_{\rm syn,max})^{1/2}$ at $\nu_{\rm syn,c}<\nu<\nu_{\rm
syn,max}$.  This implies a hard X-ray luminosity, in the 10--100keV
band, of $\sim\epsilon_e L/30$, which is comparable to $\epsilon_e
L/\Lambda$.

By specifying the acceleration mechanism, we may derive an estimate
for $h\nu_{\rm syn,min}$. If the wind power is carried by the kinetic
energy of the outflowing plasma and the plasma is heated through
internal shocks (which also accelerate particles) within the outflow,
then the characteristic temperature of the protons is
$\sim\epsilon_pm_pc^2$. This follows from the fact that relative
motions within the plasma rest frame are expected to be mildly, but
not highly, relativistic. Consider, for example, two equal mass
elements moving along the same directions with Lorentz factors
$\Gamma_1\gg\Gamma_2\gg1$. The Lorentz factor of these mass elements
in their center of mass frame is $\sqrt{\Gamma_1/\Gamma_2}/2$,
implying that a mildly relativistic relative motion is obtained unless
their respective Lorentz factors are very different.  Thus, if
electrons are coupled to the protons and carry a fraction $\epsilon_e$
of the energy density, then the lowest energy electrons will carry an
energy of $E_{e,\rm min}\sim(\epsilon_e/\epsilon_p) m_p c^2$, giving
\begin{equation}\label{eq:nu_m}
    h\nu_{\rm syn,min}\approx0.02(\epsilon_e/\epsilon_p)^2
    (\epsilon_B L_{47})^{1/2}\Gamma^{-2}\delta t_{\rm yr}^{-1}\,{\rm eV},
\end{equation}
and
\begin{equation}\label{eq:nuc_num}
    \frac{\nu_{\rm syn,c}}{\nu_{\rm
    syn,min}}<3\times10^{-5}f^{-4}\Gamma^6E_{20}^{-4}\delta t_{\rm yr}^2.
\end{equation}
Using Eq.~(\ref{eq:Gamma}) we also have
\begin{equation}\label{eq:nu_m_max}
    h\nu_{\rm syn,min}<0.01(\epsilon_e/\epsilon_p)^2
    (\epsilon_B L_{47})^{3/10}E_{20}^{-4/5}\delta t_{\rm yr}^{-3/5}\,{\rm eV}.
\end{equation}

The emission at a photon energy $\gg10$~MeV is dominated by inverse
Compton (IC) up-scattering of synchrotron photons. For collisionless
shock acceleration in internal shocks with $E_{e,\rm
min}\sim(\epsilon_e/\epsilon_p) m_p c^2$, the gamma-ray luminosity at
photon energies $\gg10$~MeV is $\nu
L_\nu=\min[1,\epsilon_e/\epsilon_B]\epsilon_e L/\Lambda$. If $E_{e,\rm
min}\gg(\epsilon_e/\epsilon_p) m_p c^2$, the IC emission may be
limited to photon energies $\gg100$~MeV, and may be shifted beyond the
observable range ($>0.1$~TeV). For $\epsilon_e\gg\epsilon_B$, the
synchrotron luminosity is suppressed to $\nu L_\nu=\epsilon_e
(\epsilon_B/\epsilon_e)^{1/2}L/\Lambda$, modifying the factor
$(\epsilon_e/\epsilon_B)$ in Eq.~(\ref{eq:Lmin}) to
$(\epsilon_e/\epsilon_B)^{1/2}$ and the factor
$(\epsilon_e/\epsilon_p)$ in Eq.~(\ref{eq:n_phot_flares}) to
$(\epsilon_e\epsilon_B)^{1/2}/\epsilon_p$.

The high energy, $\gtrsim 100$~MeV, emission may be suppressed by pair
production. A photon of high energy $E_\gamma\gg m_ec^2$ may interact
with lower energy photons, $E'_\gamma\sim\Gamma^2(m_ec^2)^2/E_\gamma$,
to produce e$^{+}$e$^{-}$ pairs. The optical depth is
$\tau_{\gamma\gamma}=n_\gamma(E'_\gamma)\sigma_{\gamma\gamma}r/\Gamma$,
where $n_\gamma(E'_\gamma)$ is the co-moving number density of photons
with observed energy $E'_\gamma$, and $\sigma_{\gamma\gamma}$ is the
e$^{+}$e$^{-}$ annihilation cross-section.  Using
$n_\gamma(E'_\gamma)\approx \nu L_\nu/4\pi r^2c \Gamma E'_\gamma$ and
the lower limit on $\Gamma$ (Eq.~\ref{eq:Gamma}), we find
\begin{equation}\label{eq:tau}
    \tau_{\gamma\gamma}\lesssim 10^{-3} \frac{(\nu
    L_\nu)_{46}}{E_{20}^{12/5}(\epsilon_B L_{47})^{3/5}} \delta t_{\rm
    yr}^{1/5}\left(\frac{E_\gamma}{m_ec^2}\right).
\end{equation}
We therefore conclude that pair production may suppress the $100$~MeV
flux only for $(\nu L_\nu)>10^{47}{\rm erg~s^{-1}}$.

\section{Luminosity function constraints}
\label{sec:LFs}

\begin{figure}
\includegraphics[width=9cm]{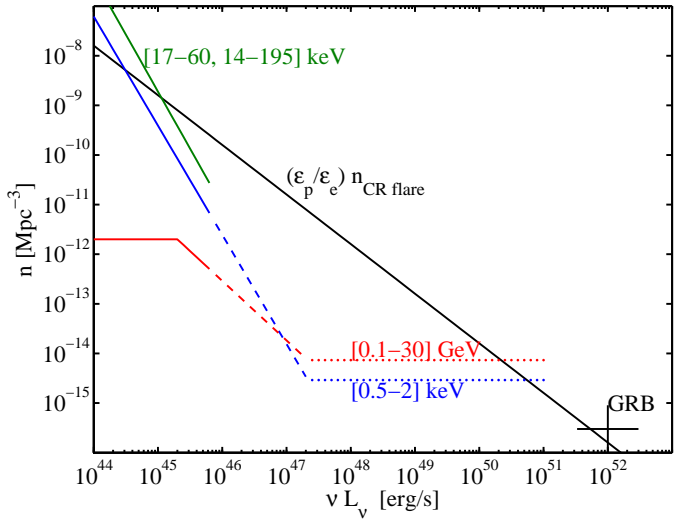}\\
\caption{The number density of active flares, $n_{CR,flare}\equiv
\dot{n}\Delta t$, required to account for the observed flux of UHECRs
(Eq.~\ref{eq:n_phot_flares}), compared to the cumulative number
density of bright extra-Galactic sources at various energy bands:
0.5-2~keV \citep[ROSAT,][]{Miyaji00}, 17-60~keV
\cite[INTEGRAL,][]{Sazonov07}, 15-195~keV \citep[Swift
BAT,][]{Tueller08}, and $>100$~MeV \cite[EGRET,][]{Chiang1998}.  The
measured luminosity in different bands is converted to $\nu L_\nu$
assuming a photon index of -2 (consistent with the observed
spectra). The solid segments of the curves represent the measured
component of the local ($z\lesssim0.2$) luminosity function (LF),
whereas the dashed segments of the curves represent the LF component
which is inferred by measuring the number density of bright sources at
higher redshift and then evolving it to $z\sim 0$ using the LF
evolution with $z$ as measured at lower $\nu L_\nu$. The dotted
segments of the curves represent the upper limit on the number density
in the luminosity range where no sources have been observed. The GRB
number density and luminosity \citep[e.g.][]{Guetta05} are shown for
comparison.}\label{fig1}
\end{figure}

Equation~(\ref{eq:n_phot_flares}) provides the number density of
active flares required to account for the observed flux of UHECRs as a
function of flare luminosity.  Figure~\ref{fig1} compares this result
with the cumulative number density of high luminosity AGN at $z<0.2$
in the energy bands of 0.5-2~keV \citep[ROSAT,][]{Miyaji00}, 17-60~keV
\cite[INTEGRAL,][]{Sazonov07}, 15-195~keV \citep[Swift
BAT,][]{Tueller08}, and $>100$~MeV \cite[EGRET,][]{Chiang1998}. At the
high luminosities under consideration, all the soft X-ray (0.5--2~keV)
sources in the ROSAT survey are identified (including stars and X-ray
clusters in addition to AGN), and all but one of the Swift BAT sources
are identified\footnote{The identification of BAT sources is more
complete than that of INTEGRAL sources because of the accurate
positions provided by the Swift XRT follow-up.}. We also note that
obscuration by a high column density of hydrogen of soft X-ray sources
can not dramatically alter the source number density since the AGNs
selected in the hard or soft X-ray bands have similar number densities
\citep[see also][]{Silverman07}.  At the highest energy band,
$>100$~MeV, where the angular resolution is poorest, source
identification is incomplete, but the contribution of unidentified
sources can not lead to a significant change in the statistics of
sources. In particular, EGRET had detected 60 high-latitude point
sources that have not been identified, compared with the 44
high-latitude sources identified as AGN \citep{Chiang1995}. The hard
X-ray (17-60~keV, -195~keV) luminosity function (LF) shown in Fig. 1
is given by
\begin{equation}\label{eq:hXLF}
    n_{hX}=10^{-11}(\nu L_\nu)_{46}^{-2.2}\,\rm Mpc^{-3}.
\end{equation}

It is important to emphasize that the luminosity function constraints
depicted in Fig. 1 refer to all the known bright sources on the sky,
and so our conclusions do not apply exclusively to AGN flares, but to
any other class of flaring sources.

We first consider the case of near equipartition between electrons and
magnetic fields, $\epsilon_e/\epsilon_B\sim1$. In this case, the flare
luminosity should be $\nu L_\nu\gtrsim10^{46}{\rm erg~s^{-1}}$ (see
Eq.~\ref{eq:Lmin}). It is obvious from Fig. \ref{fig1}, or from
comparing Eqs.~(\ref{eq:n_phot_flares}) and~(\ref{eq:hXLF}), that for
$\epsilon_e/\epsilon_B\sim1$ the observed number density of
sufficiently bright sources is much smaller than the density of active
flares required to account for the UHECR flux, unless
$\epsilon_e/\epsilon_p\ll1$. For $\nu L_\nu\gtrsim10^{47}{\rm
erg~s^{-1}}$ the number density of active flares is limited to
$<10^{-14}{\rm Mpc}^{-3}$, which implies based on
Eq.~(\ref{eq:n_phot_flares}) that
$\epsilon_p/\epsilon_e\gtrsim10^3(\nu L_\nu)_{47}^{-1}$, and
$\epsilon_p L=\Lambda(\nu L_\nu)(\epsilon_p/\epsilon_e)
\gtrsim10^{51}\Lambda_1{\rm erg~s^{-1}}$. For $\nu
L_\nu\sim10^{46}{\rm erg~s^{-1}}$, the number density of sources
appears to be consistent with the required number density of active
flares in Eq.~(\ref{eq:n_phot_flares}) for
$\epsilon_e/\epsilon_p\sim0.1$. However, the X-ray sources identified
could be candidate UHECR sources only if they are transient, and only
a small fraction of the sources observed are variable. \citet{Grupe01}
examined 113 bright ROSAT AGN on a time scale of $\sim6$~yrs, and
found that only 3 showed a factor of $10$ variation over this time
scale (all others varied by a factor less than 3). Similarly, Winter
et al. (2008) compared {\it XMM-Newton} and {\it Swift XRT}
observations of 17 sources and found fractional variations of only a
few tens of percent over $\sim 100$ days (see their Table 12),
suggesting that $\lesssim 3\%$ of all hard X-ray sources have a
lifetime of $\lesssim 10$ years.  This implies that the number density
of X-ray sources variable on $\sim5$~min (the typical integration time
in the analysis of Grupe et al. 2001 is $\sim300$~s) to $\sim10$~yr
time scale is
\begin{equation}\label{eq:hXLFvar}
    n_{hX,\rm var}\sim3\times10^{-13}(\nu L_\nu)_{46}^{-2.2}\,\rm Mpc^{-3}.
\end{equation}
Comparing with Eq.~(\ref{eq:n_phot_flares}), this implies that UHECR
flares must satisfy $\epsilon_p/\epsilon_e>500$ and $\epsilon_p
L\gtrsim10^{50}\Lambda_1{\rm erg~s^{-1}}$ for $(\nu L_\nu)_{46}=1$.  A
similar constraint is obtained using EGRET's LF.

The requirement $\epsilon_p L\gtrsim10^{50}\Lambda_1{\rm erg~s^{-1}}$
may be avoided if the magnetic field energy density is much higher
than the electron energy density, $\epsilon_e/\epsilon_B\ll1$. For
$\epsilon_e/\epsilon_B<10^{-2}$, the minimum flare luminosity is
(Eq.~\ref{eq:Lmin}) $\nu L_\nu<10^{44}{\rm erg~s^{-1}}$, and the
required number density of active X-ray flares in
Eq.~(\ref{eq:n_phot_flares}) is consistent with the number density of
variable X-ray sources in Eq.~(\ref{eq:hXLFvar}) for
$\epsilon_e/\epsilon_p\sim1$. The gamma-ray luminosity is suppressed
by $\epsilon_e/\epsilon_B$ with $\nu L_\nu<10^{42}{\rm erg~s^{-1}}$, a
range in which the number density of sources is poorly constrained by
EGRET.  Next, we consider the $\epsilon_e/\epsilon_B\sim 0.1$
regime. Here the minimum flare luminosity is $\nu L_\nu\sim10^{45}{\rm
erg~s^{-1}}$ and the required number density of active X-ray flares is
consistent with the number density of variable X-ray sources for
$\epsilon_e/\epsilon_p<10^{-2}$ and with EGRET's LF for
$\epsilon_e/\epsilon_p<10^{-3}$ (see Fig.~\ref{fig1}).  As mentioned
in \S~\ref{sec:flares}, IC emission may be shifted above the
observable range ($>0.1$~TeV) in electromagnetically-dominated
outflows. For $\epsilon_e/\epsilon_B\sim 0.1$ and
$\epsilon_e/\epsilon_p<10^{-2}$ we get $\epsilon_p/\epsilon_B>10$,
which implies that the outflow can not be electromagnetically
dominated. This, in turn, implies that the flares should be
accompanied by observable gamma-ray emission, and hence that
$\epsilon_e/\epsilon_p<10^{-3}$ must be satisfied.

The preceding discussion implies that for a flare duration in the
range $1{\rm hr}\lesssim\Delta t\lesssim10$~yr, the requirement
$\epsilon_p L\gtrsim10^{50}\Lambda_1{\rm erg~s^{-1}}$ may be avoided
only for $\epsilon_e/\epsilon_B<10^{-2}$ or
$\epsilon_e/\epsilon_p<10^{-3}$.  These constraints are likely to
improve in the near future with new $\gamma$-ray data from the
recently launched {\it GLAST}
satellite\footnote{http://glast.gsfc.nasa.gov/}, and with proposed
X-ray telescopes such as {\it
EXIST}\footnote{http://exist.gsfc.nasa.gov/}.  Next, we consider the
case of flares with $\Delta t\gg10$~yr. Since there is little
information on source variability on such time scales, all observed
sources are flare candidates. For $\Delta t\gtrsim 100$~yr, the active
flare number density is $>10^{-9}{\rm Mpc}^{-3}$ (see
Eq.~\ref{eq:n_lim}). At this density, the X-ray LF requires (see
Fig.~\ref{fig1}) an X-ray flux $\nu L_\nu<10^{45}{\rm erg~s^{-1}}$,
which implies through Eq.~\ref{eq:Lmin} that
$\epsilon_e/\epsilon_B<0.1$. The EGRET LF requires either
$\epsilon_e/\epsilon_p<10^{-3}$ or that the IC gamma-ray emission be
shifted outside the observable range, which may be possible for flares
that are electromagnetically dominated.

There is one important caveat to the above constraints. The required
number density of sources is low, $\lesssim 0.1{\rm Gpc^{-3}}$, so
that that no source is expected to be detected within a distance of
$\sim1$~Gpc, which is the GZK horizon of particles with
$E\sim10^{19}$~eV. This implies that snapshot surveys can not provide
useful constraints on the local ($z\sim 0$) number density of high
luminosity flares.  For this reason, the $z\sim 0$ LFs shown in
Fig.~\ref{fig1} are not measured directly at high luminosities, $\nu
L_\nu>10^{45.5}{\rm erg~s^{-1}}$. Rather, the number density of bright
sources is measured at a higher redshift and the local number density
is inferred from the evolution of the LF with $z$ as measured at lower
values of $\nu L_\nu$. For example, the number density of soft X-ray
sources with $\nu L_\nu>10^{46}{\rm erg~s^{-1}}$ is measured to be
$\sim 5\times10^{-11}{\rm Mpc}^{-3}$ at $z\sim1$ and inferred (but not
measured) to be much lower than $\sim10^{-12}{\rm Mpc}^{-3}$ at $z\sim
0$, based on the LF evolution measured at lower $\nu L_\nu$
\citep[see, e.g. Figure 5 of][]{Hasinger05}. Long-term monitoring
surveys offer much better prospects for constraining the source
population than snapshot surveys. For example, if the flare duration
is a few days, then a survey that lasts for a year can put constraints
that are $\sim 100$ times better than a snapshot survey. Upcoming
surveys, such as Pan
STARRS\footnote{http://pan-starrs.ifa.hawaii.edu/} are expected to
provide relevant data soon, and planned surveys such as
LSST\footnote{http://www.lsst.org/} will provide better constraining
power in the future.

We can not exclude the possibility that the number density of flaring
sources with $\nu L_\nu>10^{46}{\rm erg~s^{-1}}$ does not decrease
towards $z\sim 0$ as fast as the number density of lower $\nu L_\nu$
sources, and remains at a level of $\sim 5\times10^{-11}{\rm
Mpc}^{-3}$, which is marginally consistent with that required for the
local production rate per unit volume of UHECRs. However, such a
scenario is unnatural since it requires two coincidences: the flares
must become a dominant source of energy output only at $\nu
L_\nu>10^{46}{\rm erg~s^{-1}}$ (or else they would modify the observed
LF evolution at lower $\nu L_\nu$), and exist only at $z\sim 0$ (or
else we would observe them at $z\gtrsim 0.5$).

\section{"Dark, proton-only" flares}
\label{sec:hidden}

It is difficult to rule out a scenario in which the UHECR flares
involve "electromagnetically-dark" or "proton only" flares. Although
there is currently no evidence or physical reasoning to motivate the
consideration of a new class of hidden sources, we nevertheless
discuss its required properties for the sake of generality.

Since the cross-section for inelastic $pp$ collisions is much smaller
than the Thomson cross-section, $\sigma_T=6.7\times 10^{-25}~{\rm
cm^{2}}$, the X-ray emission may be suppressed (without affecting
proton escape from the source) by postulating the UHECR source to be
embedded within an opaque plasma cloud of column density, $\gtrsim
\sigma_T^{-1}\sim10^{24}{\rm cm}^{-2}$, which is optically thick to
Compton scattering. If the outflow is jet-like and relativistic,
scattering within the cloud would suppress the X-ray luminosity by a
factor $>\Gamma^2/\theta_j^2$, where we assume that the jet opening
angle $\theta_j>1/\Gamma$. A suppression of the expected X-ray
luminosity, $\gtrsim10^{47}~{\rm erg~s^{-1}}$, by a large factor,
$>\Gamma^4$, would allow a sufficiently high number density of
candidate UHECR sources to satisfy current limits on their
electromagnetic luminosity.

Similarly, since the cross-section for $p\gamma$ (pion
photo-production) collisions is much smaller than the Thomson
cross-section, the gamma-ray emission may be suppressed (without
affecting proton escape from the source) by postulating the source to
be embedded within an isotropic X-ray radiation field, with
sufficiently high photon density to prevent the escape of gamma-rays
through pair-production. The required photon column density,
$\sim10^{25}{\rm cm}^{-2}$, implies an X-ray luminosity of $\sim
6\times10^{43}(R/10^{16}{\rm cm})(E_\gamma/1{\rm keV}){\rm erg~s^{-1}}$,
where $R$ is the source size and $E_\gamma$ is the energy of the
background photons.

The predicted X-ray emission could also be suppressed by assuming that
the source is embedded in an intense isotropic radiation field
at IR, optical or UV frequencies, with an energy density far
exceeding that of the magnetic field of the outflow, thus suppressing
the synchrotron emission of the electrons by rapid IC cooling. For a
relativistic outflow, the luminosity $L_{\rm iso}$ associated with the
isotropic radiation field could be much smaller than that associated
with the outflowing magnetic field, $\epsilon_B L$, since the energy
density ratio in the plasma rest frame is $\Gamma^4 L_{\rm
iso}/\epsilon_B L$.  Note that such X-ray suppression does not change
the conclusion of the second paragraph of \S~\ref{sec:LFs}, that
$\epsilon_pL>10^{50}{\rm erg~s^{-1}}$ is required for flares with
X-ray luminosity of $\nu L_\nu>10^{46}{\rm erg~s^{-1}}$.  The flux can
be written as $\nu L_\nu=(\epsilon_e/f_X) L/\Lambda$ with
a suppression factor $f_X>1$,
implying that Eq.~(\ref{eq:n_phot_flares}) should be modified to
\begin{equation}\label{eq:n_phot_flares_fX}
    \dot{n}\Delta t=\frac{\epsilon_e}{2\epsilon_p f_X}\frac{\dot{\varepsilon}}
    {\nu L_\nu} =1.6\times10^{-10}\frac{\epsilon_e}{f_X\epsilon_p}
    \frac{\dot{\varepsilon}_{44}}{(\nu L_\nu)_{46}} {\rm Mpc}^{-3}.
\end{equation}
However, since the relation between proton and photon luminosities is
also modified to $\epsilon_pL=\Lambda(\nu
L_\nu)(f_X\epsilon_p/\epsilon_e)$, the constraint on $\epsilon_pL$ is
independent of $f_X$. The X-ray suppression may affect, however, the
constraint $\epsilon_e/\epsilon_B\ll1$, that must be satisfied in the
absence of X-ray suppression for flares with $\nu L_\nu\ll10^{46}{\rm
erg~s^{-1}}$ (Eq.~\ref{eq:Lmin}). In the presence of X-ray
suppression, we may write Eq.~(\ref{eq:Lmin}) as $\Gamma^2<0.1(\nu
L_\nu)_{45}f_X(\epsilon_e/\epsilon_B)^{-1}$. Combined with
$f_X\sim\Gamma^4L_{\rm iso}/\epsilon_BL<1(\Gamma/10)^2L_{\rm iso,45}$
(see Eq.~\ref{eq:L_Bmin}), this implies $L_{\rm
iso}>10^{48}(\epsilon_e/\epsilon_B) (\nu L_\nu)_{45}^{-1}{\rm
erg~s^{-1}}$. This isotropic luminosity requires the associated AGN to
involve the most massive black holes in the Universe ($\sim
10^{10}M_\odot$) shining near their limiting (Eddington) luminosity,
in order for the X-ray suppression to have a significant effect on our
results. The absence of known sources of this extreme luminosity
within the GZK horizon of UHECRs in the local Universe (see Fig. 11b
in Greene \& Ho 2007, and Fig. 6 in Hopkins et al. 2007) rules out
long-lived sources, but still allows for rare flares.  We note that
the minimum flaring time associated with the light crossing-time of
the Schwarzscild radius of these black holes is $\gtrsim 1~{\rm day}$,
and that the IR-UV variability of active AGN is observationally
constrained to be weak on longer timescales \citep{Sesar1,Sesar2} of
up to several decades \citep{Vries}.

\section{Conclusions}
\label{sec:conclusions}

The absence of steady sources of sufficient power to accelerate UHECRs
within the GZK horizon of 100~Mpc, implies that UHECR sources are
transient.  We have shown that UHECR "flares" should be accompanied by
strong X-ray and $\gamma$-ray emission. Figure 1 demonstrates that
X-ray and $\gamma$-ray surveys constrain flares which last longer than
$\sim5$~min and less than a decade to satisfy at least one of the
following conditions: {\it (i)} $L>10^{50}{\rm erg~s^{-1}}$; {\it
(ii)} the power carried by accelerated electrons is lower by a factor
$>10^2$ than the magnetic field power or by $>10^3$ than that carried
by accelerated protons; or {\it (iii)} the sources exist only at low
redshifts, $z\ll1$. The implausibility of requirements {\it (ii)} and
{\it (iii)} argue in favor of transient sources with $L>10^{50}{\rm
erg~s^{-1}}$.  The required luminosity is well above the brightest
luminosity ever recorded in an AGN flare, and exceeds by two orders of
magnitude the Eddington limit for a black hole of $10^{10}M_\odot$,
the highest mass expected to exist within a distance of 100Mpc
\citep{Lauer,Natarajan}. The results shown in Fig. 1 exclude the
regime of low flare luminosities considered by Farrar \& Gruzinov
(2008).

The lower bound of $\sim 300$s on the window of flare durations over
which our constraints apply, originates from the integration time of
the X-ray data used in Fig. 1.  For flare durations $\Delta t\lesssim
300$s, Eq. (\ref{eq:L_Bmin1}) requires $\epsilon_BL>0.3\times
10^{50}~{\rm erg~s^{-1}}$, not significantly different from the
minimum luminosity inferred for longer flare durations.

We have also explored potential caveats to the above conclusions.
Long-duration ($\gtrsim 100$ years) flares which are
electromagnetically dominated could evade the constraints illustrated
in Fig. 1 if their gamma-ray emission peaks outside the EGRET energy
band. In addition, an unknown population of
``electromagnetically-dark'' flares is in principle possible (see \S
\ref{sec:hidden} for details), although there is no physical
motivation to make its existence natural.

Future gamma-ray observations with {\it GLAST} 
and X-ray observations with {\it EXIST} would improve the
statistical constraints on the source population of UHECRs and
potentially shed more light on their nature.

\acknowledgements

We thank Glennys Farrar for useful discussions. AL thanks the Einstein
Minerva Center at the Weizmann Institute for its hospitality during
the inception of this work. This work was supported in part by NASA
grants NNX08AL43G and LA.

\end{document}